\documentclass[prd,onecolumn,aps,amsmath,amssymb,nofootinbib,
preprintnumbers,superscriptaddress]{revtex4}

\IfFileExists{srcltx.sty}{\usepackage[active]{srcltx}} 

\usepackage{subfigure}
\usepackage[dvipdfm]{graphicx}
\usepackage{amssymb}
\usepackage{amsmath}
\usepackage{amsfonts}

\newcommand{\nc}{\newcommand}

\def\ds{\displaystyle}
\nc{\be}[1]{\begin{equation}\mbox{$\label{#1}$}}
\nc{\bea}[1]{\begin{eqnarray} \mbox{$\label{#1}$}}
\newcommand{\ee}{\end{equation}}
\newcommand{\beqy}{\begin{eqnarray}}
\newcommand{\eeqy}{\end{eqnarray}}

\begin{document}

\title{Gravitational waves from the fragmentation of a supersymmetric condensate}

\author{Alexander Kusenko}
\affiliation{Department of Physics and Astronomy, University of California, Los
Angeles, CA 90095-1547, USA}
\affiliation{Institute for the Physics and Mathematics of the Universe,
University of Tokyo, Kashiwa, Chiba 277-8568, Japan}
\author{Anupam Mazumdar}
\affiliation{Physics Department, Lancaster University, Lancaster, LA1 4YB, UK}
\affiliation{Niels Bohr Institute, Blegdamsvej-17, Copenhagen, DK-2100, Denmark}
\author{Tuomas Multam\"aki}
\affiliation{Department of Physics, University of Turku, FIN-20014, Finland}

\begin{abstract}
We discuss the production of gravity waves from the fragmentation of a  supersymmetric condensate in 
the early universe.  Supersymmetry predicts the existence of flat directions in the potential.  At the end of inflation, 
the scalar fields develop large time-dependent vacuum expectation values along these flat directions.   Under 
some general conditions, the scalar condensates undergo a fragmentation into non-topological solitons, Q-balls.  
We study this process numerically and confirm the recent analytical calculations showing that it can produce gravity 
waves observable by Advanced Laser Interferometer Gravitational-Wave Observatory (LIGO), Laser Interferometer 
Space Antenna (LISA), and Big Bang Observer (BBO).  The fragmentation can generate gravity waves with an amplitude 
as large as $\Omega_{_{\rm GW}}h^2\sim 10^{-11}$ and with a peak frequency ranging from mHz to $10$~Hz, depending 
on the parameters.  The discovery of such a relic gravitational background radiation can open a new window on the physics 
at the high scales, even if supersymmetry is broken well above the electroweak scale. 
\end{abstract}

\preprint{UCLA/08/TEP/33}
\maketitle

\section{Introduction}

Supersymmetry is an appealing candidate for physics beyond the
Standard Model.  A generic feature of the scalar potential in supersymmetric 
generalizations of the Standard Model is the presence of  flat directions 
parameterized by some   { gauge invariant} combinations of 
squarks and sleptons.  At the end of cosmological inflation, the
formation of a scalar condensate along the flat directions can have a number of
important consequences~\cite{reviews}.  In particular, it can be responsible
for the matter-antimatter asymmetry generated via Affleck--Dine (AD) 
mechanism~\cite{AD}, dark matter in the form of 
Q-balls~\cite{dark_matter,Kusenko:2005du,KS,Enqvist:1998xd,ls}, 
squarks and sleptons driven inflation~\cite{AEGM,AKM},
and a curvaton~\cite{curvaton}. Note that for an inflation and a curvaton
mechanism to succeed the flat  direction must dominate the energy density of 
the universe until they decay into Standard Model baryons.

In general, a supersymmetric condensate is unstable. An initially homogeneous 
condensate can break up into lumps of the scalar field, called  Q-balls~\cite{Coleman}, 
under some very generic conditions~\cite{KS}.  All phenomenologically acceptable supersymmetric 
generalizations of the  Standard Model admit Q-balls~\cite{Kusenko:1997zq},
which can be stable, or can decay into fermions~\cite{Kusenko:1997zq,DKS}.
In many cases, the origin of the instability can be traced to running of the mass of the flat 
direction, due to logarithmic corrections~\cite{reviews}.  For the squark 
directions, the leading order correction is usually negative due to the negative contribution of the gaugino loops~\cite{EMc}. When the condensate oscillates, the negative mass correction gives rise to an average negative pressure, which triggers the instability in the 
condensate~\cite{EJM}. There are modes which grow exponentially, and the condensate starts fragments into lumps, i.e., Q-balls. 
There are many analytical~\cite{KS,Pawl,Johnson:2008se,Minos} and numerical~\cite{KK,MV,EJMV,Palti,Campanelli}  studies of Q-ball formation and their interactions~\cite{MV1}.

It was recently pointed out that the process of fragmentation can serve as a source of 
gravity waves~\cite{KM} with a detectable amplitude and with a peak frequency ranging 
from from 1~{\rm mHz} to $10$~{\rm Hz}.  This range of frequencies will be explored 
by a combination of upcoming detectors,  such as the Laser Interferometer Gravitational 
Wave Observatory (LIGO) \cite{LIGO}, the Laser Interferometer Space Antenna 
(LISA)~\cite{LISA}, the Big bang Observer  (BBO)~\cite{BBO} and the Einstein Telescope~\cite{EINSTEIN}. The spectrum of 
gravity waves is peaked near the longest wavelength, of the order of the fragmented region or of the
Q-ball size (corrected for the red shift). We will discuss how the fragmentation of the 
condensate yields gravity waves  and we will present both analytical and numerical results.


\section{Flat directions and Q-balls}

When the Standard Model is augmented by the scalar fields that carry baryon and lepton 
numbers, the non-topological solitons, or Q-balls, appear in the spectrum of such a 
theory~\cite{Kusenko:1997zq}.  At the end of inflation, large Q-balls, whose vacuum expectation values (VEV) are 
aligned with the flat direction, can form by fragmentation of the AD condensate~\cite{KS,reviews}.
The largest amplitude of gravity waves is attained when the flat direction condensate density is 
comparable to the total the energy density~\cite{KM}. Usually this is not the case in AD baryogenesis, because 
the baryon and or lepton number carried by the condensate is constrained by the present-day baryon 
asymmetry of the universe.   However, some flat directions, namely those with $B-L=0$, are not constrained 
because the net $(B+L)$ asymmetry is destroyed by the electroweak sphalerons.  The baryon 
and lepton number violating operators can  contribute to the destruction of the Q-balls and can prevent 
them from dominating the energy density of the universe.

The Q-ball with a global charge $Q$ has the following properties.  The scalar field inside the 
Q-ball has the  form 
\begin{equation}
\Phi(x,t)=\phi(x) \exp(i \omega t),
\end{equation}
where  $\phi$ is real, and $\omega\sim m_{3/2} \sim 0.1-10$~TeV in  gravity  mediated  
supersymmetry breaking models.  The global charge of the Q-ball is given by 
\begin{equation}
Q= \omega \int dx \, \phi^2(x).
\end{equation}
As long as the field has a time-dependent phase, it is associated with a non-zero global charge.

In addition, there are flat directions with both $B=0$ and $L=0$.   This means that, in 
the $\{{\rm Re}\, \Phi, {\rm Im}\, \Phi \}$ plane, the field undergoes radial motion without phase rotations.  
This can happen in the case of a flat direction inflaton~\cite{AEGM}, or in the case of a flat 
direction curvaton~\cite{curvaton}.  Due to the lack of a net charge, the fragmentation process 
generically leads to Q-balls and anti-Q-balls~\cite{EJMV,KEM0,KEM}, which eventually decay.  Part of the 
condensate energy  goes directly into exciting the gauge bosons and gauginos, which eventually thermalize the universe.   
In what follows we will refer to \textit{rotations} and \textit{radial oscillations} of the flat directions in 
the $\{{\rm Re}\, \Phi, {\rm Im}\, \Phi \}$ plane, depending on whether the flat direction carries a net global charge.  

The supersymmetric flat directions, by virtue of their couplings to the 
Standard Model fields, receive radiative corrections \cite{reviews}.  These 
corrections depend on the type of supersymmetry breaking.  A typical potential contains soft 
supersymmetry breaking mass term and higher order non-renormalizable terms which arises 
by integrating out the heavy fields above the cut-off scale $M$\footnote{For a gauge mediated case the potential
along the flat direction is different, as discussed below. However, the energy density of 
the condensate is usually much lower than that in the gravity-mediated case, hence the 
fragmentation in the gauge-mediated case does not generate a comparable amount of gravity waves.}:
\begin{equation}
    \label{qpot}
    V = m_{3/2}^2 |\Phi|^2
    \left[ 1 + K\log\left(\frac{|\Phi|^2}{M^2}\right)\right ]+ Am_{3/2}\left(\frac{\Phi^{d}}{dM^{d-3}}+h.c.\right)
    +\frac{|\Phi|^{2d-2}}{M^{2d-6}} \,.
\end{equation}
Here we have included the baryon and lepton number violating operators that are essential for AD baryogenesis 
and which play a role in decay of Q-balls, as discussed below.  
We consider $M\sim M_{Pl}\sim 2.4\times 10^{18}$~GeV. It can be seen from the 
analyses of Ref.~\cite{Gherghetta:1995dv,DRT_AD,DRT} that most flat directions in MSSM are lifted 
by monomials of dimension $4$.  The soft 
supersymmetry breaking mass term,  $m$, is proportional to
the gravitino mass, i.e. $m\sim m_{3/2}\sim {\cal O}(100)$~GeV-${\cal O}(1)$~TeV in gravity mediated supersymmetry 
breaking scenarios. The coefficient $K$ is a parameter which depends on the flat direction,
and the logarithmic contribution parameterizes the 
running of the flat direction potential.  The value of $K$ can be computed 
from the Renormalization Group (RG) equations, which, to one loop,  
give negative corrections due to gaugino loops~\cite{reviews}:  
\begin{equation}
    K \sim -\frac{\alpha}{8\pi}
        \frac{m_{1/2}^2}{m_{\tilde{\ell}}^2}\,,
\end{equation}
where $m_{1/2}$ is the gaugino mass and $m_{\tilde{\ell}}$ is the
slepton mass\footnote{A potential of the form of eq.~(\ref{qpot}), with a negative value of $K$, can be obtained  
for a generic inflaton potential which has couplings to fermions and bosons, where the fermions belong to  
a larger representation than the bosons.  The value of $K$ is determined by the Yukawa
interaction, $h\Phi \bar\psi\psi$~\cite{KEM};
$K\sim C{h^2}/{16\pi^2}$,
where $C$ is the number of fermionic loops and $h$ is the Yukawa coupling. }.
In the next section we will show that, whenever the mass of the homogeneous condensate receives a
negative correction, the condensate undergoes fragmentation, an instability which leads to formation
of Q-balls and, possibly, anti-Q-balls~\cite{KEM,reviews}.


\section{Amplification of fluctuations}

Let us consider the growth of the fluctuations in the flat direction condensate, which can be rotating~\cite{reviews,KS} or oscillating~\cite{AEGM,AKM,curvaton}.  The fluctuations in the field tend to grow when the average pressure is negative~\cite{KS,dark_matter,Kusenko:1997zq,EMc,EJM,KK,MV}, which can be the case for $K<0$ in eq.~(\ref{qpot}).  For field values $\phi \ll M$, we find:
\begin{equation}
    \label{pot0}
    V(\phi) \simeq \frac{1}{2}m_{3/2}^2\phi^2
    \left(\frac{\phi^2}{2M^2}\right)^K \propto \phi^{2+2K}\,.
\end{equation}
where we assume $|K|\ll 1$.  The equation of state for a field rotating in such a potential is 
\begin{equation}
    \label{state}
    p\simeq \frac{K}{2+K} \rho \simeq -\frac{|K|}{2}\rho \,,
\end{equation}
where $p$ and $\rho$ is a pressure and energy density of the scalar
field, respectively.  Evidently, the negative value of $K$ corresponds to the 
negative pressure, which signals the instability of the condensate. A linear perturbation
analysis~\cite{KS,KEM}  shows that the fluctuations grow exponentially if the following condition
is satisfied (see Appendix 1):
\begin{equation}
    \frac{k^2}{a^2}\left( \frac{k^2}{a^2}+2m_{3/2}^2K \right) < 0.
\end{equation}
Clearly, the instability band exists for negative $K$, as expected from the negative pressure arguments~\cite{reviews}. The 
instability band, $k$,  is in the range~\cite{KS,dark_matter,EMc,KK,KEM}
\begin{equation}
    \label{k-band}
    0 < \frac{k^2}{a^2} < \frac{k_{max}^2}{a^2} \equiv  2m_{3/2}^2|K|\,,
\end{equation}
where $a$ is the expansion factor of the universe. 
The most amplified mode lies in the middle of the band, and the maximum growth rate  of the perturbations, see 
eq.~(\ref{linear-growth1}) in Appendix 1,  is determined by  
$\dot{\alpha} \sim ~|K|m_{3/2}/2$~\cite{KEM}. 
The initial growth of perturbations can be described analytically in the linear regime by   eqs.~(\ref{linear-growth1},\ref{linear-growth2}) with 
$\alpha(t)\sim |K|m_{3/2}\Delta t/2$.  When $\delta \phi/\phi_0\sim {\cal O}(1)$, the  fluctuations become nonlinear.  
This is the time when the homogeneous condensate breaks down into Q-ball.  We will study this process numerically.

%
%

The energy density in the condensate depends on the model, and, foremost, 
on the type of supersymmetry breaking terms that lift the flat direction.
This is because the potential along the flat direction depends on supersymmetry
breaking (it vanishes in the limit of exact supersymmetry), and there are many
ways to break supersymmetry. In the gauge-mediated supersymmetry breaking scenarios
the potential can have the form~\cite{KS}
\begin{equation}
V(\phi)\approx m_{S}^4\log\left(1+\frac{|\phi|^2}{m_{S}^2}\right)\,.
\label{gaugemediatedsusybreaking}
\end{equation}
Here $m_S$ is the scale of supersymmetry breaking, which is of the order of
1~TeV.

The main difference between gauge and gravity mediated cases for us is the mass
per charge stored in the flat direction condensate and in the Q-balls that form
eventually as a result of the fragmentation. In the gravity mediated scenarios,
the mass density is $ \rho_0 \sim m_{3/2}^2 \phi_0^2 $, the global charge
density is $ n_Q\sim m_{3/2} \phi_0^2$, and the mass per unit global charge is of
the order of $m_\phi$, independent of the VEV $\phi_0$, for a review see~\cite{reviews}. In gauge mediated
scenarios, where the Q-ball radius $R\sim m_S^{-1} Q^{1/4}$, $\omega \sim m_S Q^{1/4}$, $\phi_0\sim m_S Q^{1/4}$, 
the mass density is $ \rho_0 \sim m_S^4$,  the global charge
density is $ n_Q\sim \omega \phi_0^2$, and the mass per unit global charge is
$\rho_0/n_Q\sim m_S^2/\phi_0$~\cite{KS,DKS}.  Furthermore, in a gauge mediated scenario the
flat direction condensate never dominates the energy density of the universe.   In what follows, we will 
concentrate on gravity mediated supersymmetry breaking scenarios.


\section{Gravity waves}

The gravity waves are generated because the process of fragmentation involves inhomogeneous, non-spherical, 
anisotropic motions of the scalar condensate. As a result, the stress energy tensor receives anisotropic 
stress-energy contribution. The fragmentation of the condensate\footnote{Generation of gravity
wave from the coherent oscillations of the inflaton field has been studied in 
Refs.~\cite{GarciaBellido:2007af,Dufaux,Easther}.} takes place on spatial scales smaller than  the Hubble 
radius. The gravity waves are generated at the time of fragmentation, i.e. when the linear perturbation in the flat direction
condensate starts growing, as in eqs.~(\ref{linear-growth1},\ref{linear-growth2}).  

In calculating energy density of  the gravitational waves, we follow 
the transverse-traceless (TT) components of the 
stress-energy momentum tensor. By perturbing the 
Einstein's equation, we obtain the evolution of the tensor perturbations~\cite{GarciaBellido:2007af}:
\begin{equation}
\ddot h_{ij}+3H\dot h_{ij}-\frac{\nabla^2}{a^2}h_{ij}=
16\pi G \Pi_{ij}\,,
\end{equation}
where $\partial_{i}\Pi_{ij}=\Pi_{ii}=0$ and 
$\partial_{i}h_{ij}=h_{ii}=0$.  The TT part of the spatial
components of a symmetric anisotropic stress-tensor
$T_{\mu\nu}$ can be found by using the spatial projection operators,
$P_{ij}=\delta_{ij}-\hat{k}_i\hat{k}_j$, with $\hat k_{i}=k_{i}/k$:
\begin{equation}
\Pi_{ij}(k)=\Lambda_{ij,mn}(\hat k) T_{mn}(k)\,,
\end{equation}
where $\Lambda_{ij,mn}(\hat k)\equiv \left(P_{im}(\hat k)P_{jn}(\hat k)-(1/2)P_{ij}(\hat k)P_{mn}(\hat k)\right)$.
The TT perturbation is written as $h_{ij}(t,{\bf\hat{k}})=\Lambda_{ij,lm}(\hat{k})u_{ij}(t,\bf{k})$, where
\begin{equation}
\ddot u_{ij}+3H\dot u_{ij}-\frac{1}{a^2}\nabla^2 u_{ij}=16\pi GT_{ij}\,.\label{uequ}
\end{equation}

The source terms for the energy momentum tensor in our case are just the gradient terms of the flat direction
condensate, 
\begin{equation}
\label{ttsource}
T_{ij}=\frac{1}{a^2}(\nabla_i\phi_1\nabla_j\phi_1+\nabla_{i}\phi_2\nabla_{j}\phi_2)\,, 
\end{equation}
where $ \phi_1$ and $ \phi_2$ represent the real and imaginary parts of $\phi$, respectively. 
The gravitational wave (GW) energy density is given by
\be{rhogw}
\rho_{GW}=\frac{1}{32\pi G}\frac{1}{V}\int d^3\vec{k}\, \dot{h}_{ij}\dot{h}^*_{ij},
\ee
where $V$ is the volume of the lattice. To estimate the magnitude of gravitational wave energy density on a lattice, we 
approximate eq. (\ref{rhogw}) by:
\be{rhogw2}
\rho_{GW}\approx\frac{1}{32\pi G}\frac{1}{V}\int d^3\vec{x}\, \dot{u}_{ij}\dot{u}^*_{ij}.
\ee
In numerical calculations we track the evolution of $\Omega_{GW}$ using the $u_{12}$ and $u_{21}$ components\footnote{We have verified the approximate equivalence between $u_{ij}$ and $h_{ij}$ by comparing the numerical results for the gravity waves obtained in the two approaches.  }.

\begin{figure}[ht!]
\subfigure[$m_{3/2}t=0$]{\includegraphics[]{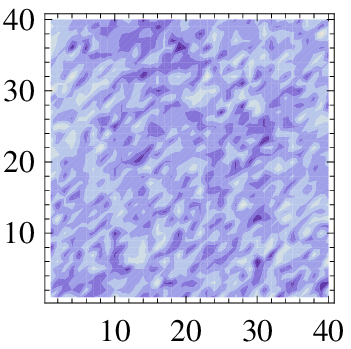}}
\subfigure[$m_{3/2}t=75$]{\includegraphics[]{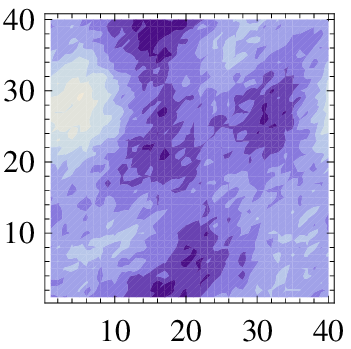}}
\subfigure[$m_{3/2}t=150$]{\includegraphics[]{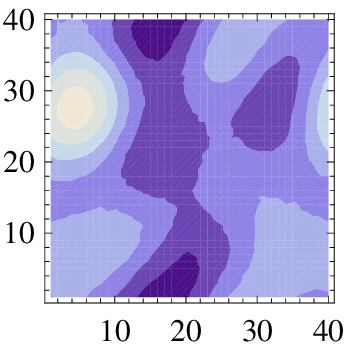}}
\subfigure[$m_{3/2}t=375$]{\includegraphics[]{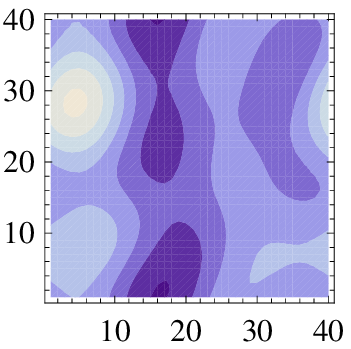}}
\subfigure[$m_{3/2}t=525$]{\includegraphics[]{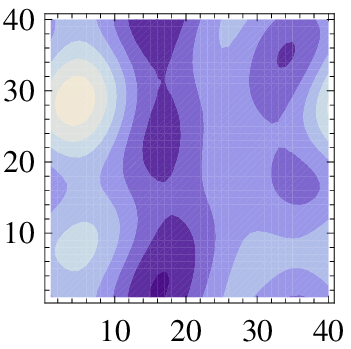}}
\subfigure[$m_{3/2}t=675$]{\includegraphics[]{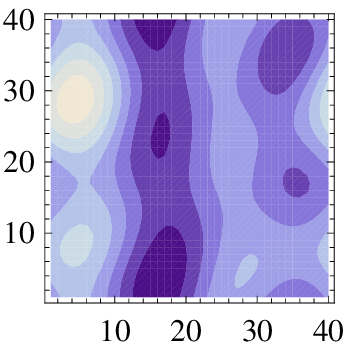}}
\subfigure[$m_{3/2}t=825$]{\includegraphics[]{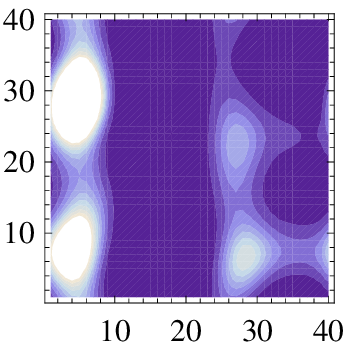}}
\subfigure[$m_{3/2}t=900$]{\includegraphics[]{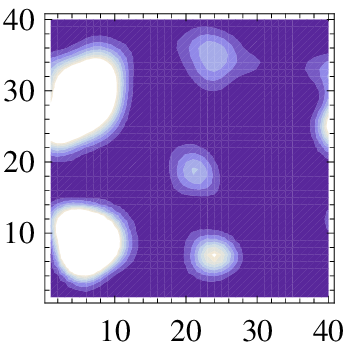}}
\caption{Evolution of the condensate in two dimensions, calculated on the $N=64$ lattice.
White areas are the instability regions which eventually form Q-balls. The fragmentation occurs on the time scale 
$t\sim {\cal O}(10^2-10^{3}) m_{3/2}^{-1} $.}\label{slices}
\end{figure}

\begin{figure}[ht!]
\subfigure[$m_{3/2}t=900$]{\includegraphics[height=4cm]{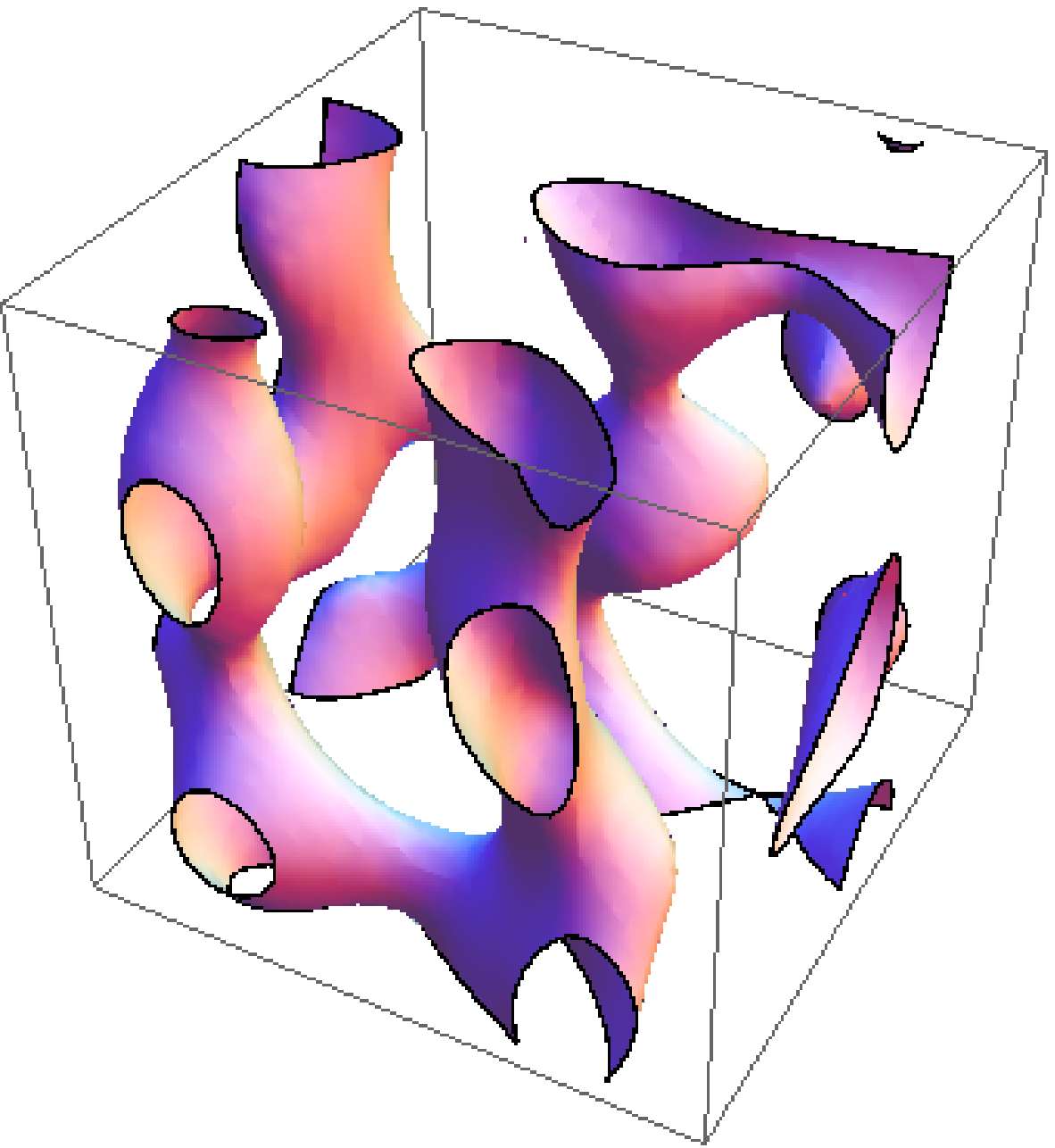}}
\subfigure[$m_{3/2}t=1050$]{\includegraphics[height=4cm]{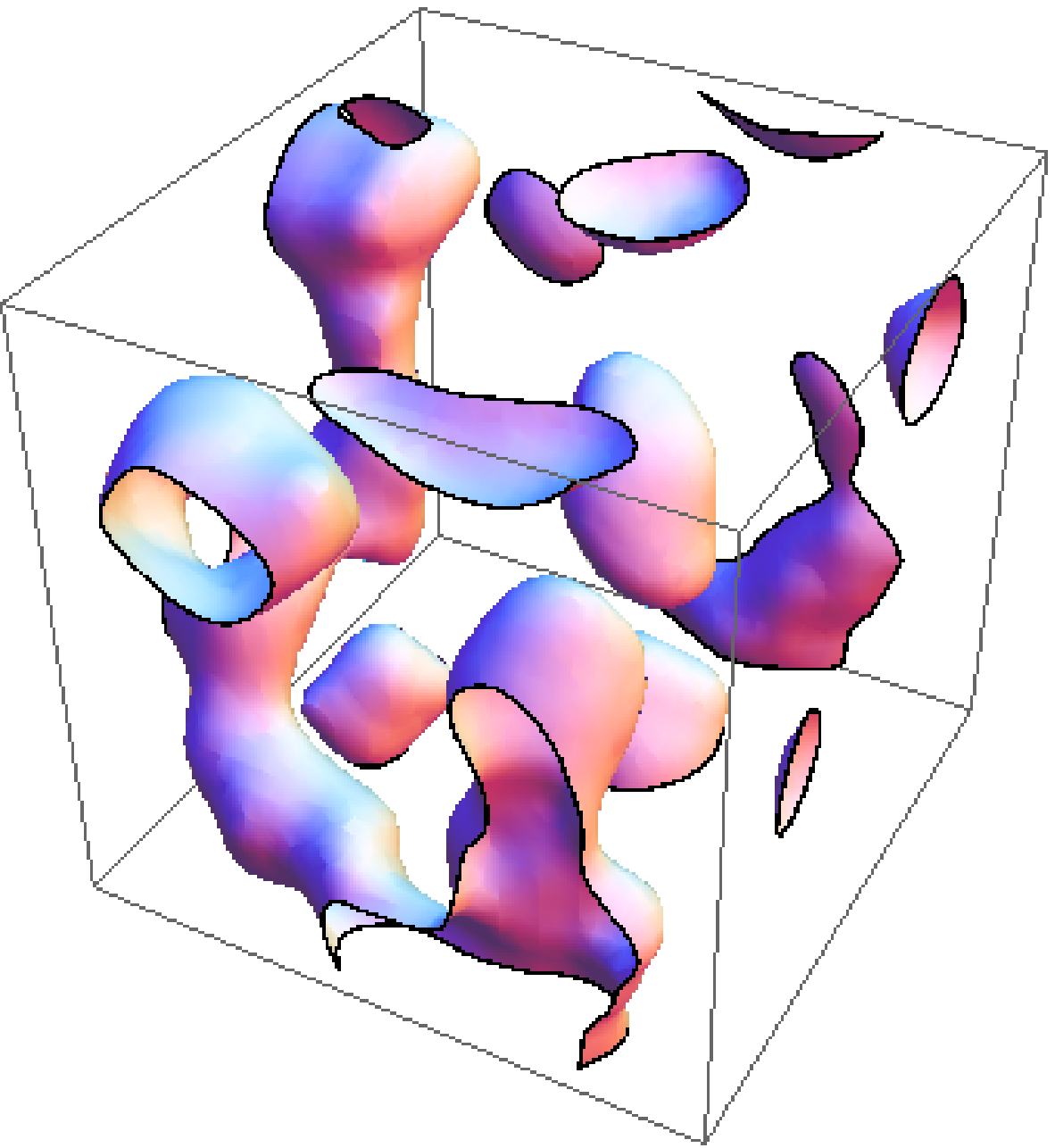}}
\subfigure[$m_{3/2}t=1200$]{\includegraphics[height=4cm]{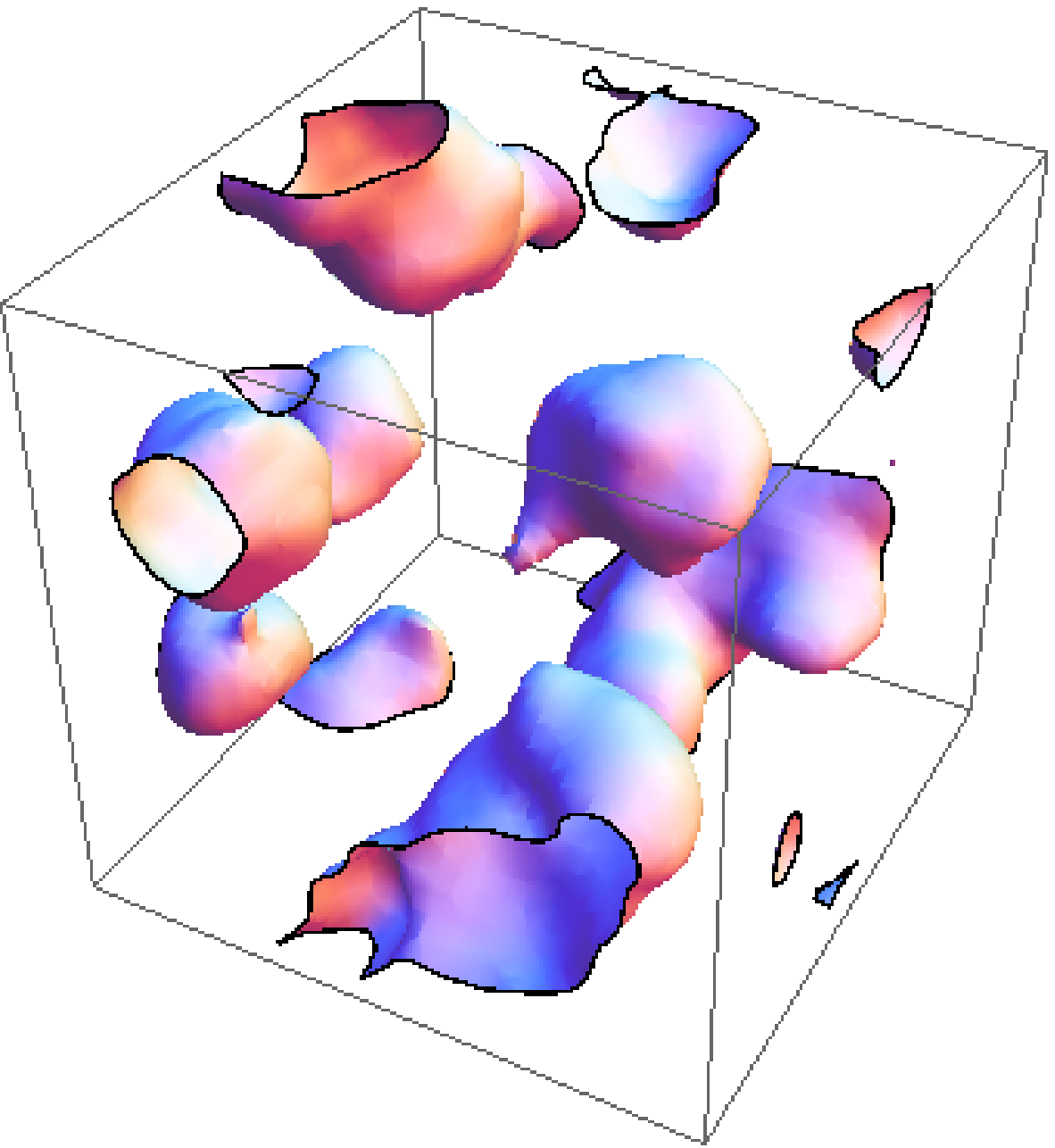}}
\subfigure[$m_{3/2}t=1350$]{\includegraphics[height=4cm]{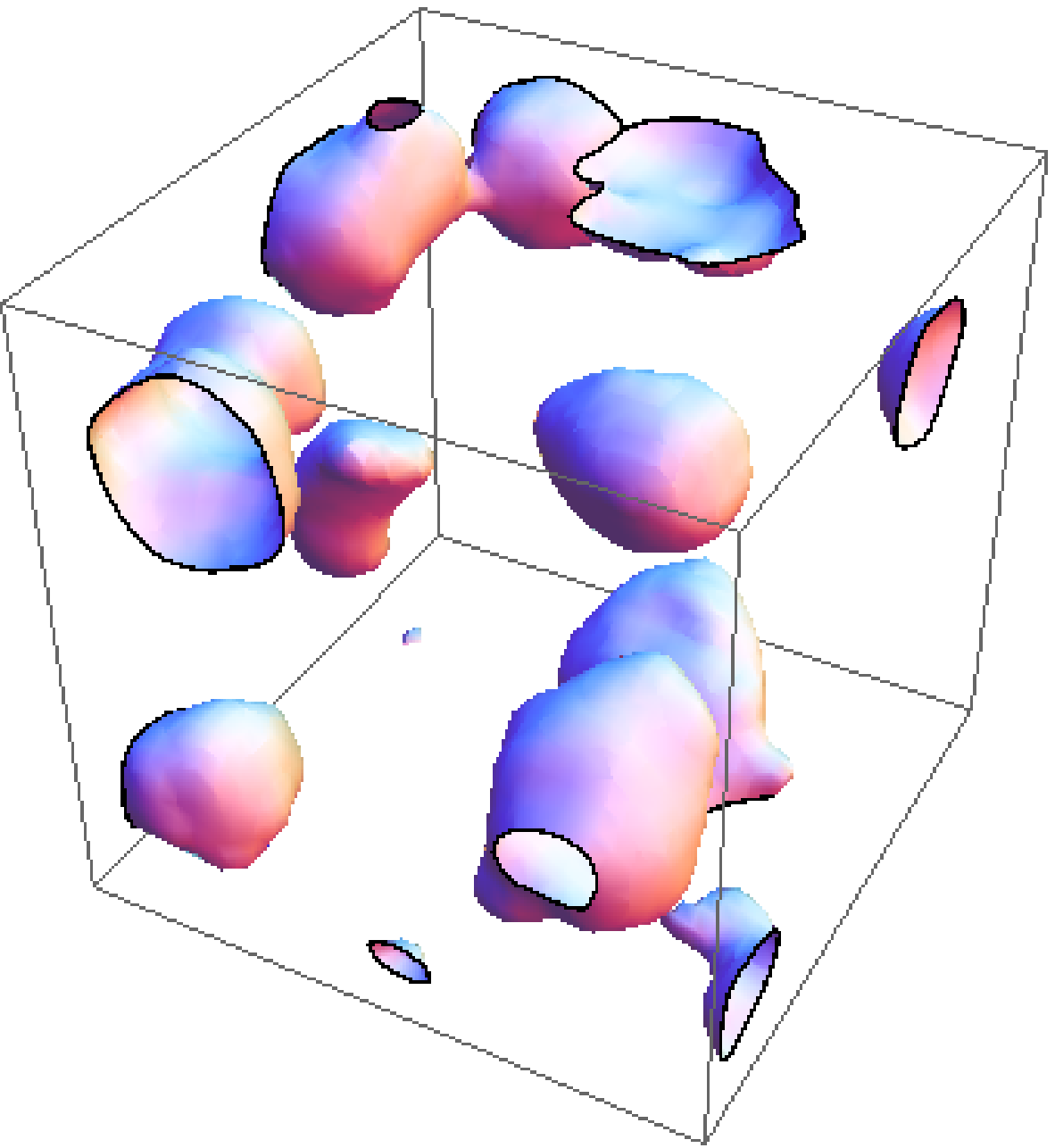}}
\caption{Three-dimensional evolution of the condensate on the $N=64$ lattice. Lumps of Q-matter are shown 
in color. The Q-matter  breaks up into isolated lumps, Q-balls.}\label{boxes}
\end{figure}

\section{Analytical approximations}

Let us estimate the fraction of energy density stored in the gravitational waves produced 
by the fragmentation of the condensate.  The best-amplified mode 
is $k^2=k_{max}^2 \approx m_{3/2}^2 |K|(1-|K|/4)$ and the maximal growth rate of the fluctuation 
$\delta\phi=\delta\phi_0 e^{\alpha(t)+ikx}$  is $\dot{\alpha}=|K|m_{3/2}/2$.  Therefore we obtain:
\be{flucs1}
\dot{(\delta\phi)}=\dot{\alpha}\delta\phi=\frac 12 |K| m_{3/2}\delta\phi\,.
\ee
During the fragmentation process the field perturbation grows from the initial value $\delta\phi_0$: 
$$\delta\phi(t,x)=\delta\phi_0\exp(m_{3/2}|K|t/2+ikx).$$ 
Using the above approximation in eq. (\ref{uequ}), we obtain 
\be{uest}
m_{3/2}^2\ddot u-k_{max}^2u\approx \frac{2}{M_{Pl}^2}k_{max}^2(\delta\phi(t,x))^2 .
\ee
Therefore, the fastest growing mode can be approximated by: 
$$u(t)\approx 2M_{Pl}^{-2}(\delta\phi (t,x))^2.$$

The fragmentation becomes non-linear, and Q-balls form, when $\delta\phi/\phi\sim {\cal O}(1)$.
The majority of gravity waves are produced over the time interval determined by the condition; 
$\Delta t \sim 2/(|K|)\ln(\phi/\delta\phi)$. The final (saturated) gravity wave energy density can 
therefore be estimated by using eq. (\ref{rhogw2}) as:
\be{rhogwest}
\rho_{GW}\sim \frac{1}{4}M_{Pl}^2\left(\frac{du}{dt}\right)^2\sim \frac{|K|^2m^2_{3/2}\phi(t)^4}
{M_{Pl}^2}
\ee
The fractional energy density is then given by:
\be{omest}
\Omega_{GW}=\frac{\rho_{GW}}{m_{3/2}^2\phi(t)^2}\sim |K|^2\left(\frac{\phi(t)}{M_{Pl}}\right)^2\,.
\ee
Note that for a condensate, which is dominating the energy density of the universe at the time of fragmentation,
the Hubble expansion rate is very small, i.e. $m_{3/2}\gg H(t)$, see~\cite{AEGM,AKM}. For physically motivating
parameters, we have chosen; $m_{3/2}\sim 100$~GeV, $\phi(t)\sim 10^{16}$~GeV, and $H(t)\sim 1$~GeV, therefore,
for a reasonable value of $K\sim 0.1$, we obtain, $\Omega_{GW}\sim 10^{-6}$. Note that $\Omega_{GW}$ depends 
on the value of $K$, for $K=0$ there are no excitations of gravity waves.


\section{The fate of Q-balls}
\label{fate}

The formation of Q-balls is usually considered in the context of Affleck--Dine baryogenesis~\cite{AD,reviews,KS}.  In this case,  the energy density stored in the scalar condensate is small because it is related to the small baryon asymmetry of the universe.  However, there is no reason why the supersymmetric scalar condensates could not have a much higher density if they carried a zero $(B-L)$ charge, and if the Q-balls, formed via fragmentation, decayed before they came to dominate the energy density.  The former requirement is sufficient for the primordial condensate to be independent from the baryon asymmetry of the universe because the net $(B+L)$ global charge is erased by the sphalerons in the course of the electroweak phase transition.   The latter has to do with the fact that, if the Q-balls forming from the fragmentation of the scalar condensate are long-lived, they can come to dominate the energy density in the  universe causing an epoch of matter-dominated expansion that may not allow the efficient reheating at the end~\cite{Kasuya:2007cy}.  In the MSSM, there are flat directions that have $B-L=0$, for example, $QQQL$,  $ \tilde u\tilde u\tilde d\tilde e$, $QQ\tilde u \tilde d$, $QL\tilde u\tilde e$, etc. 

To estimate the range of the lifetimes of Q-balls, one must consider several decay modes.  First, the scalar fields can evaporate into fermions carrying the same global quantum numbers.  The decay of Q-balls via evaporation~\cite{Q-decay}, as well as Q-ball melting at finite temperature~\cite{ls}, are both suppressed by the surface-to-volume ratio and can lead to some very long decay times~\cite{Kasuya:2001hg,Kasuya:2007cy}.  However, in the presence of baryon and lepton number violating operators, the decay may proceed much faster, because the Q-ball is only as stable as the U(1) symmetry is good. 

Let us first consider higher-dimensional operators suppressed by the scale $M\sim M_{\rm Pl}$.   The supersymmetry preserving baryon number violating operators are given by F terms with dimensions larger than equal to $5$ and D terms with dimension larger than $6$.  For example, the following baryon and lepton number violating operators can be written as F-terms:
 \begin{equation}
 {\cal L}\supset \frac{1}{M}Q_{i}Q_{j}Q_{k}L_{i}|_{\theta^2}+\frac{1}{M}\tilde u_{i}\tilde e_{j}\tilde u_{k}\tilde d_{l}|_{\theta^2}+{\rm h.c.}\,,
  \end{equation}
 where $i,j,k,l$ are the generations ($i\neq k$). These interactions cause the $B,L$ violation by $\Delta (B-L)=0$ and $\Delta (B+L)=-2$.  A $(B+L)$-ball can lose its $(B+L)$ charge and disintegrate via  $2\leftrightarrow3$  processes that have cross section of the order of 
 $\sigma \sim {1}/{M^2}$.  
 The decay rate is given by:
 \begin{equation}
 \Gamma \sim \frac{1}{Q}\frac{dQ}{dt}\sim \sigma n_{\phi}\, 
 \sim  \gamma^2|K|^2\left(\frac{\phi_0}{M_{Pl}}\right)^2m_{3/2}\, , 
\label{decay_perturbative}
 \end{equation}
 where $n_{\phi}\sim \gamma^2|K|^2m_{3/2}\phi_0^2$ is the number density of Q-quanta inside the Q-ball. 
Although the flat direction undergoes fragmentation, but not all the number density of $\phi$ field goes into 
forming a Q-ball. Here we have provided a conservative estimation, note that for $K=0$, the Q-balls do not form.
The factor $\gamma\sim 0.1$ denotes the formation time scale of Q-balls, which is roughly given by 
$t^{-1}\sim \gamma |K|m_{3/2}$, numerically one can see from Figs.~(\ref{slices},\ref{boxes}) that 
$m_{3/2}t\sim {\cal O}(100)$ for $|K|\sim 0.1$. The initial VEV of the flat direction is $\phi_0\sim 10^{16}$~GeV 
and $M\sim M_{Pl}$. The Q-balls decay when $\Gamma^{-1}$ is of the order of the Hubble time, we find
 that the Q-balls  decay at temperature $T\sim 10^{5}$~GeV for mass  $m_{3/2}\sim 10^{2}$~GeV.

However, the estimate of decay time in equation~(\ref{decay_perturbative}) is based on simplifying assumption of  incoherent particle interactions inside the Q-ball, which may be inapplicable to scalars in a coherent state, such as Q-ball.  It may be more appropriate to treat  the Q-ball decay semi-classically, as discussed by Kawasaki et al.~\cite{Kawasaki:2005xc}.  Kawasaki et al. considered baryon number violating operators in eq.~(\ref{qpot}) that arise from supersymmetry breaking terms: $ A m_{3/2} \left ( \frac{\Phi^d}{d\, M} + {\rm h.c.}\right)$
Such operators are essential for the Affleck--Dine baryogenesis to work.  
They found numerically that, for $d=4$,   the time scale of the Q-ball decay is~ \cite{Kawasaki:2005xc}:
\begin{equation}
\Gamma \sim 10^{-5}\left(\frac{|K|}{0.1}\right)^{3/2}m_{3/2}\,,
\end{equation}
This time scale is comparable to the Hubble time when the plasma temperature is $T\sim 10^{7}$~GeV for $m_{3/2}\sim 100$~GeV and $K\sim 0.1$.  The Q-ball formation occurs when the energy density is of the order of $\left(10^8 {\rm GeV} \right)^4$, and the Hubble parameter is much larger than the decay width: $H\gg \Gamma$.  Thus, the rate of baryon number violating processes is too slow to affect the Q-ball formation, but it is fast enough for the Q-balls to decay before they can dominate the energy density of the universe.

\begin{figure}[ht!]
\includegraphics[width=10cm]{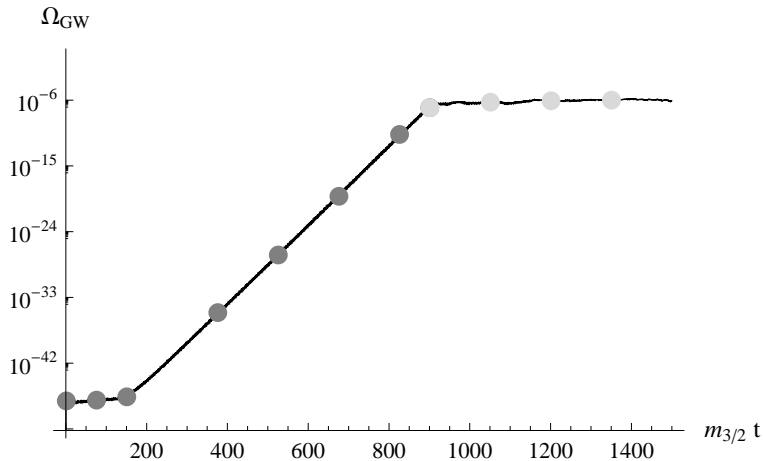}
\caption{Evolution of $\Omega_{GW}$, where the dark and light shaded dots correspond to the snapshots shown in Fig.~\ref{slices} and Fig.~\ref{boxes}, respectively.  The comparison illustrates the agreement between two-dimensional and three-dimensional calculations.  (Our final numerical results are based on the three-dimensional calculation.)}
\label{gwom}
\end{figure}


\begin{figure}[ht!]



\subfigure[$m_{3/2}t=400$]{\includegraphics[width=7cm]{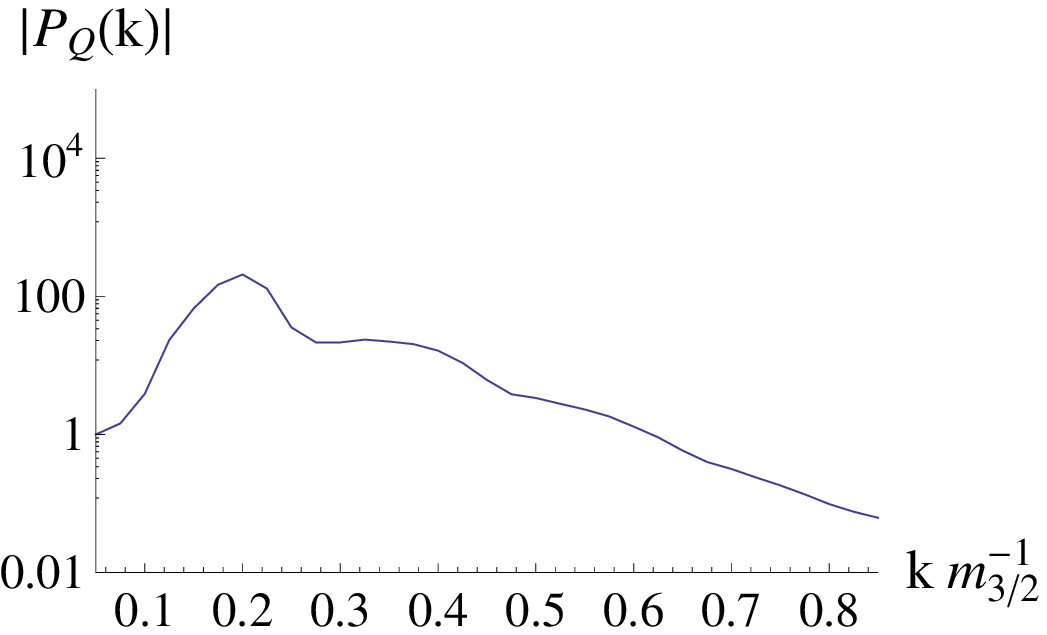}}\,~~~~~~~~~~~~~~~~~~
\subfigure[$m_{3/2}t=400$]{\includegraphics[width=7cm]{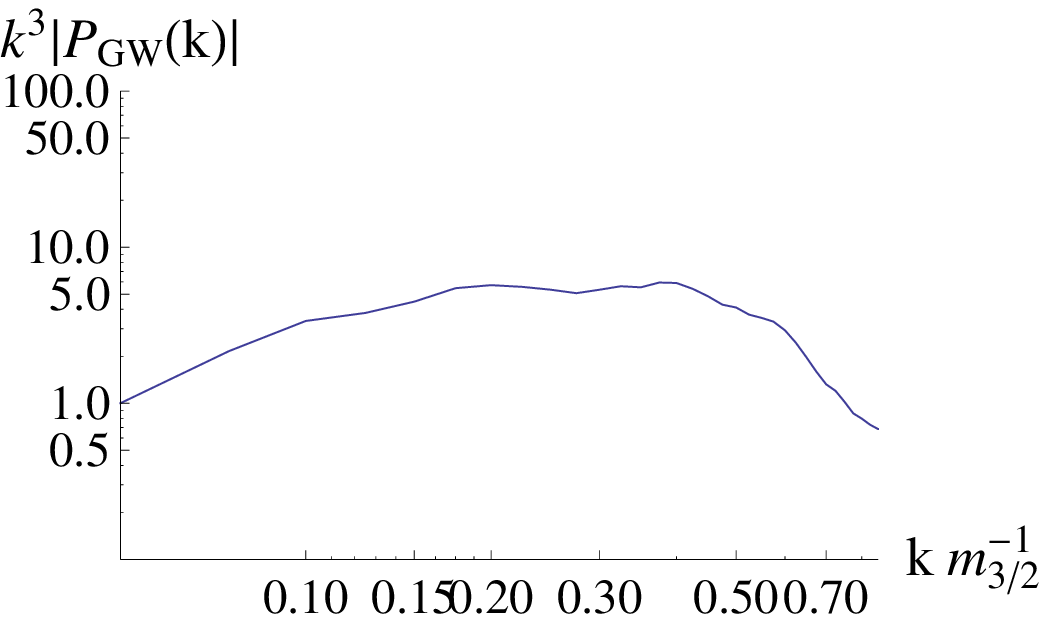}}
\caption{The panel on the left shows the spectrum of the growing perturbations in the flat direction
condensate, and the panel on the right shows the gravity wave spectrum, as a function of time, on a $N=64$ lattice. The overall normalization 
is arbitrary in both the cases.}\label{gwspecs}
\end{figure}

\begin{figure}[ht!]
\subfigure[~The final amplitude of the gravity waves does not depend on the initial perturbations.  Here 
$\phi_0=10^{16}$~GeV and $m_{3/2}=100$~GeV.]{\includegraphics[height=6cm]{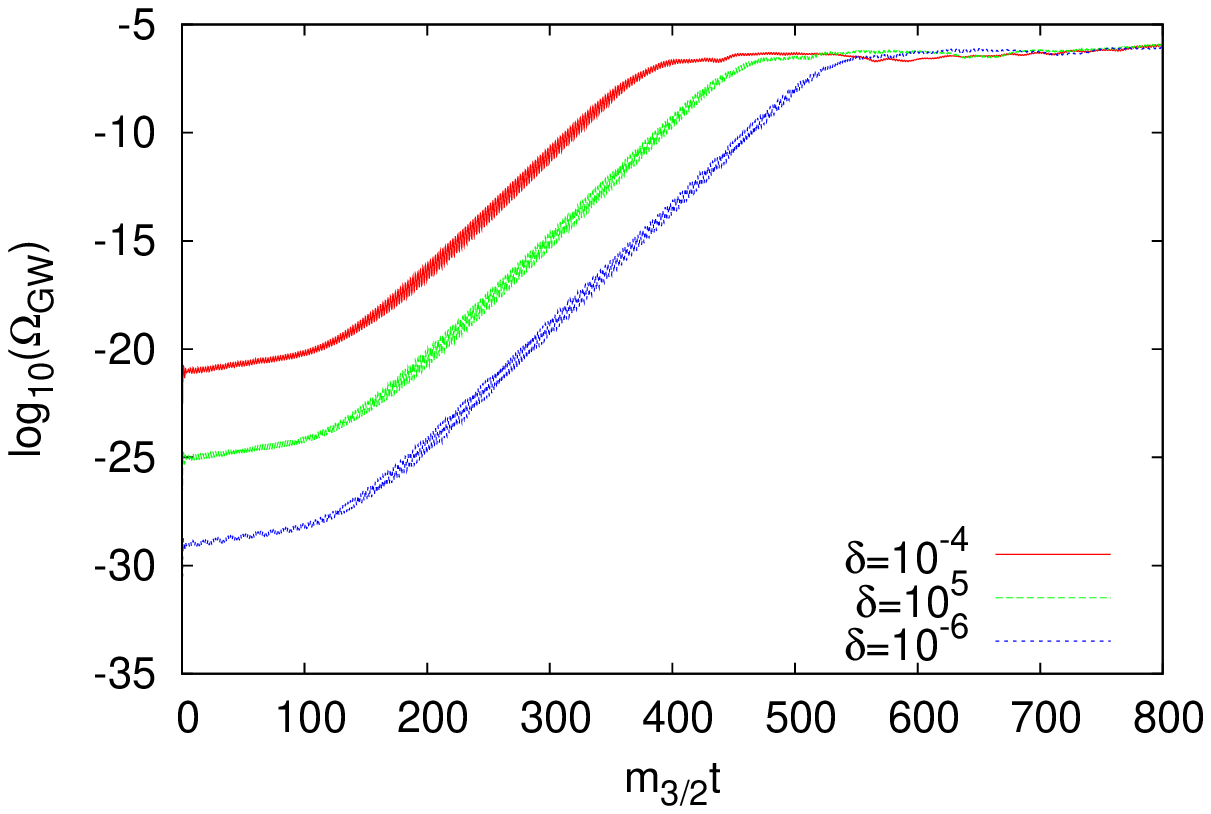}}~~
\subfigure[The final amplitude of gravity waves saturates for different values of $K$. However, 
for $K=0$, there is no fragmentation and, therefore, there is no increase in the gravity wave amplitude. The value of $|K|$ determines the growth rate of the gravity waves. Here $\phi_0=10^{16}$~GeV and $m_{3/2}=100$~GeV.]{\includegraphics[height=6cm]{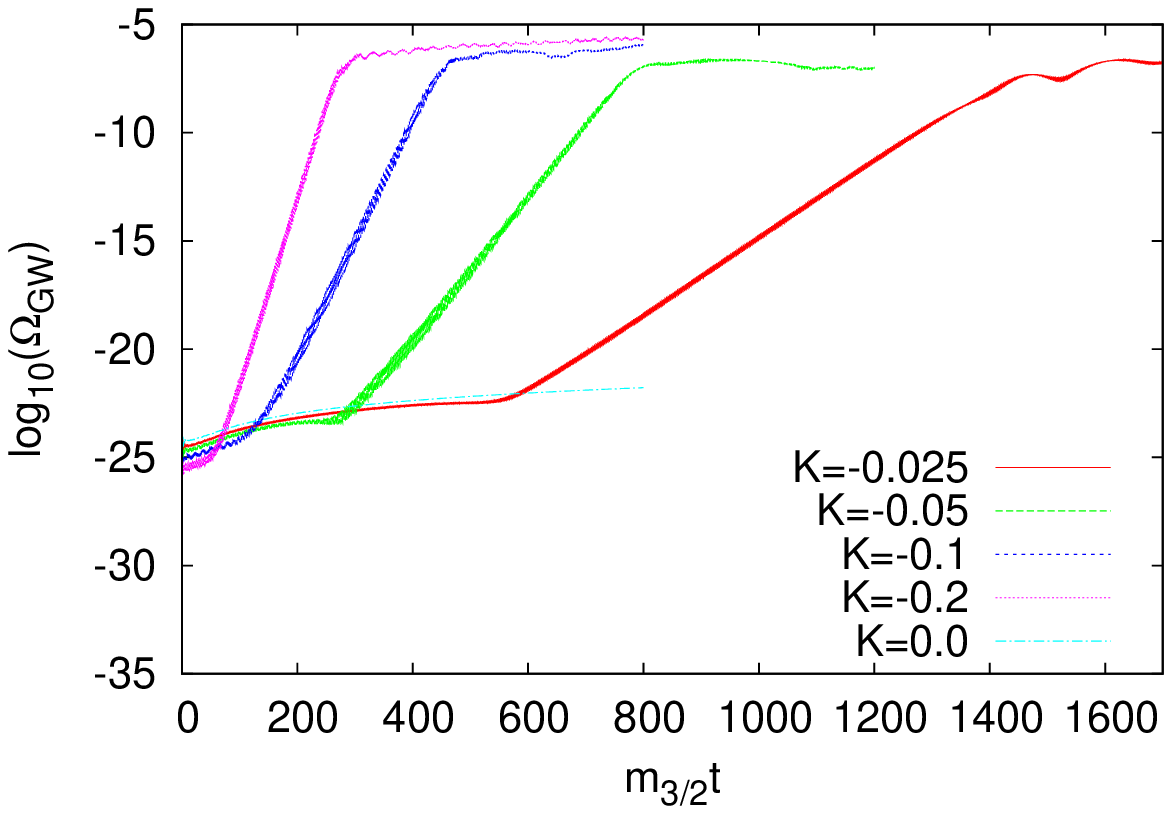}}
\caption{Effects of varying different parameters on the evolution of the gravitational wave fractional energy density $\Omega_{GW}$.}\label{rphi}
\end{figure}



\section{Numerical results}

We solved the equations of motion for the field, eqs.~(\ref{scaling-0}--\ref{numeqs}) in Appendix 2, 
numerically on a three dimensional cubic $N^3$ lattice for $N=64$, along with the evolution of  $u_{ij}$ given in 
eq.~(\ref{uequ}).  For the purpose of illustration, we have chosen the soft supersymmetry breaking mass 
$m_{3/2}=10^2$~GeV, which is motivated by gravity mediated supersymmetry breaking scenarios.
We have considered a range of values for $K$: $K\in \{0,-0.1, -0.2, -0.05, -0.025\}$\footnote{The value of $K=-0.2$ is relatively
large but we use this value for comparisons  in studying  the growth of $\Omega_{GW}$.}. The initial VEV is taken  
$\phi_0=10^{16}$~GeV. The initial condition for the time-dependent phase was set randomly, as expected at the end of inflation. 
The initial small fluctuations of the condensate are set to be $\delta \phi/\phi\sim 10^{-5}$, as expected in inflationary cosmology.

In Figs.~\ref{slices} and \ref{boxes} we show the fragmentation of the condensate.  One can see the initial snapshot of a (nearly) 
homogeneous condensate in Fig.~1a. The process of fragmentation leads to islands with growing density contrast at later times, as shown in  Fig.~1.  Similar plots with different snapshots are shown in the 3D case, in Fig.~2, where one can see the fragmentation of the condensate into Q-balls. We can see how certain modes in the band of instability begin to grow leading to an eventual fragmentation of the condensate, and how these modes stop growing. At the same time, there is mode-mode mixing which excites higher-$k$ modes, and also affects the fragmentation process as shown in  Figs.~\ref{slices}~and~\ref{boxes}. The fragmentation does not happen isotropically, and coherent, macroscopic, non-spherical motions of the condensate create a quadrupole moment which leads to the creation of gravity waves.

In Fig.~\ref{gwom} we show the evolution of the energy density stored in the gravity waves with respect to the critical energy 
density of the universe at the time of production. We plot $\Omega_{GW}$ as a function of $m_{3/2}t$ for $\phi_0=10^{16}$~GeV 
along with 2D slices and 3D isosurface plots for selected times, which  correspond to the snapshots shown in Figs.~\ref{slices} and ~\ref{boxes}. The dark grey dots in Fig.~\ref{gwom} are results of the 2D calculation, and the  light grey dots are results of the 3D calculation, at the times that correspond to the snapshots in Figs.~\ref{slices}, and~\ref{boxes}.

At the onset of fragmentation all the scalar modes are excited; however, as discussed above, only a narrow band of the initial perturbation spectrum becomes unstable. The band of instability obtained from numerical calculations is shown in Fig.~\ref{gwspecs}.  It agrees with our analytical estimate, see eq.~(\ref{k-band}). The gravitational waves spectrum is similar to that of the scalar perturbations, as shown in  Figs.~\ref{gwspecs}a,~\ref{gwspecs}b.  However the gravity wave spectra are flatter.  We have multiplied  the gravity wave power spectrum by $k^3$  to highlight the growing mode.   We have found that the peak of the gravity wave spectrum is determined by the longest wavelength which is of the order of the time scale when the Q-balls form, i.e. $m_{3/2} t\sim {\cal O}(10-100)$, see Figs.~\ref{slices} and \ref{boxes}. The growth eventually stops, once the fragmentation is near completion\footnote{With our current simulation we are unable to determine the exact peak frequency for the gravity waves. We would require a larger box size, i.e. N=256, to demonstrate this initial frequency, which is beyond the scope of the current paper. For the moment we will restrict ourselves with an approximate analytical discussion on an observable frequency, see section VIII. }. 

In Fig.~\ref{rphi}, we show the temporal evolution of the gravitational wave density for different values of the
initial field perturbations, different values of $K$ and initial angular motion. In the perturbation plot, Fig.~\ref{rphi}a, we find that the resulting gravitational wave energy density is independent of the size of the initial perturbation in the flat direction. The second plot, Fig.~\ref{rphi}b, shows different values of  $K$ determines how quickly the fragmentation proceeds, or the growth in the gravity wave density. We note that, when $K$ is set to zero, there is no growing mode, the condensate never fragments, 
and no gravity waves are produced.  



\section{ Observable signal}

After they are produced, the gravitational waves are decoupled from plasma. Let us estimate the fraction of the critical energy density $\rho_c$ stored in the gravity waves today:
\begin{eqnarray}
\label{maxamp}
\Omega_{GW} (t_0)& = &\Omega_{GW}
\left(\frac{a_{\ast}}{a_0}\right)^4\left (\frac{H_{\ast}}{H_0}\right)^2 \\ 
&\approx&  \frac{1.67\times 10^{-5}}{ h^{2}}\left(\frac{100}{g_{s,\ast}}\right)^{1/3}
\Omega_{GW}  
\approx 10^{-11} h^{-2}\, ,  \nonumber 
\end{eqnarray}
where $a_0$ and $H_0$ are the present values of the scale factor and the Hubble 
expansion rate, and $a_{\ast}$ and $H_{\ast}$ are the respective values at 
the time of the fragmentation. The estimate $\Omega_{GW}\sim 10^{-6}$
obtained from our numerical results, is within one order of magnitude of  the analytical estimates in eq.~(\ref{omest}).
LISA can detect the gravitational waves down to $\Omega_{GW}h^2\sim 10^{-11}$ at mHz frequencies.  
One can estimate the peak frequency of the gravitational radiation observed today, which is determined by the 
initial frequency, $f_{\ast}\approx \omega_{k}/2\pi$.  We obtain 
\begin{eqnarray}
\label{peak}
f&=& f_{\ast}\frac{a_{\ast}}{a_0}=f_{\ast}\left(\frac{a_{\ast}}{a_{\rm rh}}\right)
\left(\frac{g_{s,0}}{g_{s,{\rm rh}}}\right)^{1/3}
\left(\frac{T_{0}}{T_{\rm rh}} \right)\, \\
&\approx & 0.6 \, {\rm
mHz}\left(\frac{g_{s,{\rm rh}}}{100}\right)^{1/6}\left(\frac{T_{\rm rh}}
{1~{\rm TeV}}\right)\left(\frac{m_{3/2}}{t_f^{-1}}\right)\,, \nonumber 
\end{eqnarray}
where we have assumed that $a_{\ast}\approx
a_{\rm rh}$ (which also implies that one can neglect the expansion of the universe  
during the oscillations of the condensate).  The numbers of relativistic degrees of freedom are $g_{s,{\rm
rh}}\sim 300$, $g_{s,0}\sim 3.36$.  Here the subscript ``rh'' denotes the
epoch of reheating and thermalization, while the subscript ``0'' refers to the
present time.  The typical frequency of the gravity waves is be determined by 
the size of the fragmented regions. This is roughly
given by the scale at which fragmentation happens, i.e. $m_{3/2}t_{f}\sim {\cal O}(10-100)$. 
This can be seen in our numerical results shown in Fig.~\ref{slices} and Fig.~\ref{boxes}.  For
$T_{\rm rh}\sim 1$~TeV  (the value of the reheat temperature  is determined when 
the flat direction is responsible for reheating the universe~\cite{AM,AEGM}), 
the frequency is of the order of $10^{-3}-10^{-2}$~Hz,  which is in the right frequency range
for LISA~\cite{LISA}.  A higher temperature $T_{\rm rh}\sim100$~TeV corresponds 
to the frequency range, $10-100$~Hz.  
Signals in both of these ranges will be accessible to BBO~\cite{BBO} and Einstein Telescope~\cite{EINSTEIN}.  Since the supersymmetry breaking scale is related to the energy in the condensate, as well as the reheating temperature,  future gravity wave experiments
could be in a position to probe supersymmetry breaking scale above 100~TeV, which beyond the reach of the Large Hadron Collider (LHC).

The spectrum of gravity waves produced by fragmentation is expected peak near the longest wavelength, of the order of the Q-ball size, 
which is $\sim \left((0.1-0.01)m_{3/2}\right)^{-1}$.  The relatively narrow spectral width will help distinguish this signal from the gravity waves generated by inflation~\cite{inflation}, which are expected to have an approximately scale-invariant spectrum (and a smaller amplitude).  It may be difficult, however, to distinguish this signal from that generated by a  first-order phase transition~\cite{Kamionkowski:1993fg}, because both of these sources are expected to produce a relatively narrow spectrum determined by the Hubble parameter at the relevant time in the early universe.   The gravity waves could also be generated in the electroweak-scale preheating.  However, the amplitude of gravity waves expected in such a scenario~\cite{GarciaBellido:2007af}  is considerably lower than that from the fragmentation of the supersymmetric condensate.    LISA and BBO will be able to distinguish the gravity waves produced by fragmentation from those of point sources, such as merging black holes and neutron stars, which have  specific ``chirp'' properties~\cite{Owen:1998dk}.

\section{Conclusions}

We have shown that the gravitational waves of observable amplitude could be produced when a homogeneous supersymmetric flat direction condensate fragmented into small lumps, i.e., Q-balls.  The instability is a generic prediction for a supersymmetric flat direction in the early universe.  The origin of the gravity waves is in the non-spherical  anisotropic motions of the condensate that result from the growth of small initial perturbations.  

Gravity waves with the energy fraction as large as $\Omega_{\rm GW}(t_0)h^2\sim 10^{-11}$ can be generated 
with a peak frequency ranging from {\rm mHz} to $10$~{\rm Hz}, depending on the reheat temperature, which can vary in the range $1-100$~TeV. The signal in the {\rm mHz} frequency range can be detected by LISA,  while a higher  frequency $1-10$~{\rm Hz} is in the range of LIGO and BBO. The spectrum of gravity waves is different from many astrophysical sources identified by their chirp frequencies and from the gravity waves generated during inflation, which are stochastic waves with a scale invariant broad wavelength spectrum.  However, a first-order phase transition in the early universe can produce a similar spectrum of gravitational radiation~\cite{Kamionkowski:1993fg}.  

An observable amplitude of gravity waves can be produced by a class of flat directions which carry no net 
$(B-L)$ number and which are close to dominating the energy density of the universe at the time of fragmentation.  Such flat directions are not responsible for generation of the baryon asymmetry via the Affleck-Dine scenario, although they may undergo a similar cosmological evolution as the flat directions discussed in connection with the matter-antimatter asymmetry.  The  $(B+L)$ asymmetry generated by these flat directions is destroyed by the electroweak sphalerons.  Although the identification of the origin of the signal may not be unambiguous, detection of these gravity waves by LISA, BBO, or EINSTEIN could open a window on supersymmetry in the early universe even if it is realized at a very high energy scale.


\section{Acknowledgments}

AK thanks M.~Kawasaki and F.~Takahashi for helpful discussions. 
The work of AK was supported in part  by DOE grant DE-FG03-91ER40662 and by the
NASA ATFP grant  NNX08AL48G.  AM would like to thank J.~ McDonald and D.~ Lyth 
and he is partly supported by ``UNIVERSENET''(MRTN-CT-2006-035863) and  by STFC  Grant PP/D000394/1.


\section{Appendix~1}

Let us neglect any baryon number non-conservation and consider the homogeneous mode 
$\Phi = (\phi e^{i \theta})/\sqrt{2}$ that obeys the classical equations of motion and 
and fluctuations about this classical solution, 
$\phi \rightarrow \phi +\delta\phi$ and
$\theta \rightarrow \theta + \delta\theta$. The equations of
motion yield~\cite{KS,KEM}
\begin{eqnarray}
    \ddot{\phi} + 3H\dot{\phi} - \dot{\theta}^2\phi
             + V'(\phi) & = & 0, \\
    \label{theta-eq}
    \phi\ddot{\theta} + 3H\phi\dot{\theta}
      + 2\dot{\phi}\dot{\theta} & = & 0,
\end{eqnarray}
for the homogeneous mode, and
\begin{widetext}
\begin{eqnarray}
    \label{eom-fl-1}
    \delta\ddot{\phi} + 3H\delta\dot{\phi}
       - 2\dot{\theta}\phi\delta\dot{\theta}
       - \dot{\theta}^2\delta\phi  -\frac{\nabla^2}{a^2}\delta\phi
       + V''(\phi)\delta\phi & = & 0, \\
    \label{eom-fl-2}
    \phi\delta\ddot{\theta}
       + 3H\phi\delta\dot{\theta}
       + 2(\dot{\phi}\delta\dot{\theta}
             + \dot{\theta}\delta\dot{\phi})
       -2\frac{\dot{\phi}}{\phi}\dot{\theta}\delta\phi
       -\phi\frac{\nabla^2}{a^2}\delta\theta & = & 0,
\end{eqnarray}
\end{widetext}
for the fluctuations.   Furthermore, 
\begin{eqnarray}
    V'(\phi) & = & m_{3/2}^2 \phi \left[ 1 + K +
       K\log\left(\frac{\phi^2}{2M^2}\right) \right], \\
    V''(\phi) & = & m_{3/2}^2 \left[ 1 + 3K +
       K\log\left(\frac{\phi^2}{2M^2}\right) \right].
\end{eqnarray}
Due to the conservation of the global U(1) charge in the physical volume, the solution has the property $\dot{\theta}\phi^2a^3={\rm const}.$  If the energy density of the scalar field dominates the universe,
the homogeneous part of the field evolves as
\begin{eqnarray}
    \phi(t) & \simeq &
       \left( \frac{a(t)}{a_0} \right)^{-3/(2+K)} \phi_0, \\
    \dot{\theta}^2(t) &\simeq &
       \left( \frac{a(t)}{a_0} \right)^{-6K/(2+K)} m_{3/2}^2.
\end{eqnarray}

To find the most amplified mode,  we use the ansatz:
\begin{eqnarray}\label{linear-growth1}
    \delta\phi & = & \left( \frac{a(t)}{a_0} \right)^{-3/(2+K)}
      \delta\phi_0 e^{\alpha(t)+ikx}, \\
      \label{linear-growth2}
    \delta\theta & = & \left( \frac{a(t)}{a_0} \right)^{-3K/(2+K)}
      \delta\theta_0 e^{\alpha(t)+ikx}.
\end{eqnarray}
If $\dot{\alpha}$ is real and positive, these fluctuations grow exponentially, become nonlinear, and form $Q$ balls~\cite{KS}. Substituting these into eqs.(\ref{eom-fl-1}) and (\ref{eom-fl-2}), one obtains the following dispersion relation~\cite{KEM}:
\begin{widetext}
\begin{equation}
  \label{det}
   \left|
      \begin{array}{cc}
          +\dot{\alpha}^2+\frac{k^2}{a^2}+2m_{3/2}^2Ka^{-6K/(2+K)}
          & \ds{-2m_{3/2}a^{-6K/(2+K)}\phi_0
              \left(-\frac{3K}{2+K}H+\dot{\alpha}\right)} \\[3mm]
          \ds{\frac{2m_{3/2}\dot{\alpha}}{\phi_0}}
          & \ds{\ddot{\alpha}+\dot{\alpha}^2+\frac{k^2}{a^2}
            +\frac{3K}{2+K}\left[ (4-3K)H^2 -\frac{\ddot{a}}{a}
              -H\dot{\alpha}\right]}
    \end{array}
    \right| = 0,
\end{equation}
\end{widetext}
where we have set $a_0=1$. If one neglects  the cosmological expansion and assumes 
$\ddot{\alpha} \ll \dot{\alpha}^2$ for simplicity, one can reduce 
eq.(\ref{det}) to an approximate simplified equation: 
\begin{equation}
  \label{det2}
   \left|
      \begin{array}{ccc}
          \ds{\dot{\alpha}^2+\frac{k^2}{a^2}+2m_{3/2}^2K}
          & &\ds{-2m_{3/2}\phi_0\dot{\alpha}} \\[2mm]
          \ds{\frac{2m_{3/2}\dot{\alpha}}{\phi_0}}
          & & \ds{\dot{\alpha}^2+\frac{k^2}{a^2}}
    \end{array}
    \right| = 0.
\end{equation}
The perturbations grow exponentially if ${\rm Re} \, \dot{\alpha}>0$, 
which yields:
\begin{equation}
    \frac{k^2}{a^2}\left( \frac{k^2}{a^2}+2m_{3/2}^2K \right) < 0.
\end{equation}


\section{Appendix~2}

For our numerical analysis it is convenient to define dimensionless variables:
\begin{equation}\label{scaling-0}
\varphi=\phi/m_{3/2},~~~\tilde{k}=k/m_{3/2}, ~~~\tau = m_{3/2}t,~~~\xi = m_{3/2}x\,,
\end{equation}
which we will use to  study gravity waves during the non-linear fragmentation of the condensate.

Writing $\varphi = (\varphi_1 + i \varphi_2)/\sqrt{2}$, we obtain the
equations for the homogeneous mode:  
\begin{equation}
    \varphi_i''+ 3h\varphi_i' + \left[ 1 + K + K\log \left(
       \frac{\varphi_1^2+\varphi_2^2}{2M^2} \right) \right]\varphi_i
    = 0,
\end{equation}
where $h=H/m_{3/2}$, $i=1,2$, and the prime denotes the derivative with
respect to $\tau$.  For the fluctuations, one obtains 
\begin{equation}
    \left[\frac{d^2}{d\tau^2} + 3h\frac{d}{d\tau}
      + \frac{\tilde{k}^2}{a^2} + V_{ij}\right]
    \left(\begin{array}{c}
           \ds{\delta\varphi_1} \\
           \ds{\delta\varphi_2}
         \end{array} \right) = 0,
\end{equation}
where $V_{ij}$ denote the second derivatives with respect to
$\varphi_i$ and $\varphi_j$: 
\begin{eqnarray}\label{numeqs}
    & &
    V_{ii} = 1 + K + K\log \left(
              \frac{\varphi_1^2+\varphi_2^2}{2M^2} \right)
             + 2K \frac{\varphi_i^2}{\varphi_1^2+\varphi_2^2},
    \nonumber \\
    & &
    V_{12} = V_{21} = 2K \frac{\varphi_1\varphi_2}
    {\varphi_1^2+\varphi_2^2}.
\end{eqnarray}



\end{document}